\documentclass[epj,twocolumn]{svjour}  
\usepackage{amsmath,amssymb}
\usepackage{graphicx}
\usepackage{cite}
\usepackage{bm}

\journalname{Eur. Phys. J. B}
\graphicspath{{plots/}}

\newcommand{\sinc}{\mathop{\mathrm{sinc}}}
\newcommand{\NN}{\mathbb{N}}
\newcommand{\RR}{\mathbb{R}}
\newcommand{\ZZ}{\mathbb{Z}}

\newcommand{\vk}{\boldsymbol{k}}

\newcommand{\vphi}{\boldsymbol{\varphi}}

\begin{document}

\title{Generic shape of multichromatic resonance peaks}
\author{Mar{\'\i}a Laura Olivera\inst{1} \and Jes\'us Casado-Pascual\inst{1} \and Sigmund Kohler\inst{2}}
\institute{F{\'\i}sica  Te\'orica,  Universidad  de  Sevilla,  Apartado  de
Correos  1065,  41080  Sevilla,  Spain
\and
Instituto de Ciencia de Materiales de Madrid, CSIC, 28049 Madrid, Spain}

\date{\today}
\PACS{{}{}}

\abstract{
We investigate dissipative dynamical systems under the influence of an
external driving with two or more frequencies.  Our main quantities of
interest are long-time averages of expectation values which turn out to
exhibit universal features.  In particular, resonance peaks in the vicinity
of commensurable frequencies possess a generic enveloping function whose
width is inversely proportional to the averaging time.  While the universal
features can be derived analytically, the transition from the specific
short-time behavior to the long-time limit is illustrated for the examples of a
classical random walk and a dissipative two-level system both with biharmonic
driving.  In these models, the dependence of the time-averaged response on the
relative phase between the two driving frequencies changes with increasing
integration time.  For short times, it exhibits the $2\pi$ periodicity of
the dynamic equations, while in the long-time limit, the period becomes a
fraction of this value.}

\maketitle

\section{Introduction}

Dynamical systems driven by time-dependent forces represent paradigmatic
models for non-equilibrium effects in the realm of both classical and
quantum mechanics.  There one may find counter-intuitive phenomena such as
stochastic resonance \cite{Gammaitoni1998a, CasadoPascual2003a},
synchronization \cite{anishchenko:2002, freund:2003, lindner:2004,
casado1:2005, Goychuk2006a}, and the ratchet effect by which directed currents
emerge despite the absence of any net force \cite{Reimann2002a,
Hanggi2009a, CuberoRenzoni}.  The common physics behind these phenomena is
an interplay of non-linearities and non-equilibrium.  Driven systems have
been studied in the Hamiltonian limit \cite{Flach2000a, Cubero2018a,
Cubero2018b}, as well as in the steady state of dissipative classical
\cite{Reimann2002a, Hanggi2009a, CasadoPascual2013a, CasadoPascual2015a}
and quantum mechanical models \cite{Reimann1997a, Lehmann2003b,
Kohler2005a}.

Even far from equilibrium, spatio-temporal symmetries may inhibit the
emergence of a dc response such as a ratchet current.  Many of
these symmetries are based on the fact that a sinusoidal driving force
changes its sign when time is shifted by half a period.  For bichromatic or
multichromatic forces, the periods of the various forces are different and,
thus, the symmetry analysis has to be revised.  Typically, one has to
distinguish two cases, namely those of commensurable and incommensurable
frequencies.  In the former case, the symmetry depends on the phase between
the components of the driving, which has been verified with transport
experiments in quantum dots \cite{Forster2015b, Platonov2015a}.  In
reference~\cite{Forster2015b}, it has also been demonstrated theoretically and
experimentally that for incommensurable frequencies, the symmetry may be
higher, despite that the driving force is only quasi periodic and may not
possess any symmetry or anti-symmetry.

Here, we work out generic properties of dissipative dynamical systems under
multi-frequency driving, in particular the shape of the resulting resonance
peaks in the long-time limit.  Moreover, we study how this limit emerges.
The article is organized as follows.  In Section~\ref{sec:Formulation}, we
formulate the problem and in Section~\ref{sec:finitetime} derive how the
independence on the initial time provides generic properties.  In
Section~\ref{sec:bichromatic}, we consider finite-time effects for bichromatic
driving and in Sections~\ref{sec:randomwalk} and \ref{sec:tls} show how the
limits are approached for a classical random walk model and for a quantum
mechanical two-level system, respectively.  Finally, we summarize and
conclude in Section~\ref{sec:conclusions}.

\section{Formulation of the problem}
\label{sec:Formulation}

Suppose that the dynamical equations governing the time evolution of the
system under consideration depend on time through $N$ time-periodic
functions of the form
\begin{equation}
\label{eq1}
f_j(t)=\Phi_j \left(\Omega_j\,t+\varphi_j\right)\,,
\end{equation}
with $j=1,\dots,N$. In the above expression, $\Phi_j$ are $2\pi$-periodic
functions [i.e., $\Phi_j(\theta+2\pi)=\Phi_j(\theta)$ $\forall\theta\in
\mathbb{R}$] and $\Omega_j$ and $\varphi_j$ denote the angular frequency
and the phase of $f_j(t)$, respectively. The precise nature of the
functions $f_j(t)$ is irrelevant for our purposes, e.g., they may represent
external oscillatory forces, modulating amplitudes, etc. 

Assume that the system has been prepared in a state $s_0$ at
an initial time $t_0$. Depending on the case, $s_0$ may represent the
system's density operator (in the case of quantum systems), the one-time
probability density (in the case of  classical stochastic systems), the
values of a finite number of state variables or classical fields (in the
case of  classical deterministic systems), etc. For our purpose, the only
relevant fact is that all the physical properties of the system at any
subsequent time $t\geq t_0$ are uniquely determined by this initial
preparation and the dynamical equations. In particular, the (expectation)
value at time $t\geq t_0$ of a certain physical quantity $Q$ of this
system, denoted by $Q_t$, will depend on the specific values taken by the
parameters appearing in the functions $f_j(t)$, as well as, on the initial
preparation. When necessary, this dependence will be made explicit by the
notation $Q_t\left(\boldsymbol{\Omega},\vphi|s_0,t_0\right)$ with the
vector notation $\boldsymbol{\Omega}=(\Omega_1,\dots,\Omega_N)$ and
$\boldsymbol{\varphi}=(\varphi_1,\dots,\varphi_N)$.

The transformation property of $Q_t(\boldsymbol{\Omega},\vphi|s_0,t_0)$ is
determined by the $f_j$, as for any time shift $\tau$, the set of periodic
functions $\{f_j(t)\}_{j=1,\dots,N}$ is invariant under the transformation
$\{t,\vphi\}\mapsto \{t+\tau,\vphi-\boldsymbol{\Omega}\tau\}$.
Consequently, since the only explicit time-dependence of the dynamical
equations comes from the functions $f_j(t)$, it readily follows that
\begin{equation}
\label{general1}
Q_t(\boldsymbol{\Omega},\vphi|s_0,t_0)=Q_{t+\tau}(\boldsymbol{\Omega},\vphi-\boldsymbol{\Omega}\,\tau|s_0,t_0+\tau)\,.
\end{equation} 
In particular, taking $\tau=-t$, one obtains
\begin{equation}
\label{general2}
Q_t\left(\boldsymbol{\Omega},\vphi|s_0,t_0\right)=Q_{0}(\boldsymbol{\Omega},\vphi+\boldsymbol{\Omega}\,t|s_0,t_0-t)\,.
\end{equation}

Henceforth, we assume that, after a certain transient time (or, more
formally, in the limit $t_0\to -\infty$), the observable under
investigation reaches a stationary value
\begin{equation}
Q_t^{{\mathrm{st}}}(\boldsymbol{\Omega},\vphi)=\lim_{t_0\to -\infty}Q_t(\boldsymbol{\Omega},\vphi|s_0,t_0).
\end{equation} 
This assumption has been found in many dissipative systems. From
equation \eqref{general2}, it then follows that
\begin{equation}
\label{dissipative}
Q_t^{{\mathrm{st}}}(\boldsymbol{\Omega},\vphi)=Q_{0}^{{\mathrm{st}}}(\boldsymbol{\Omega},\vphi+\boldsymbol{\Omega}\,t).
\end{equation}
Therefore, in the stationary regime, the time evolution of $Q$ admits a
description in terms of a time-dependent phase vector of the form
$\vphi+\boldsymbol{\Omega}\,t$. 

Here, we are interested in the generic properties of the time average
\begin{equation}
\label{TA}
\overline{Q}_{T}(\boldsymbol{\Omega},\vphi)=\frac{1}{T}\int_0^{T} dt \,Q_t^{{\mathrm{st}}}(\boldsymbol{\Omega},\vphi),
\end{equation}
and more specifically on the infinite-time average
\begin{equation}
\label{ITA}
\overline{Q}_{\infty}(\boldsymbol{\Omega},\vphi)=\lim_{T\to \infty}\overline{Q}_{T}(\boldsymbol{\Omega},\vphi).
\end{equation}

Since the functions $\Phi_j(\Omega_jt+\varphi_j)$ are $2\pi$-periodic in
the phases $\varphi_j$, so will be $Q_t^{\mathrm{{st}}}(\boldsymbol{\Omega},\vphi)$.
Therefore, taking into account equation (\ref{dissipative}), it can be
Fourier expanded as
\begin{equation}
\label{Fourier1}
Q_t^{\mathrm{st}}(\boldsymbol{\Omega},\vphi)=\sum_{\vk\in\ZZ^N}q_{\vk}(\boldsymbol{\Omega})
e^{i \vk \cdot(\vphi+\boldsymbol{\Omega} t )},
\end{equation}
where the centered dot denotes the usual scalar product in $\RR^N$ and
\begin{equation}
\label{r2}
q_{\vk}(\boldsymbol{\Omega})=\int_{-\pi}^{\pi}\dots\int_{-\pi}^{\pi} \frac{e^{-i\vk\cdot\vphi}}{(2\pi)^N}\,Q_0^{{\mathrm{st}}}(\boldsymbol{\Omega},\vphi)d^N\vphi.
\end{equation}

By inserting equation (\ref{Fourier1}) into equation (\ref{TA}), we obtain
\begin{equation}
\label{finitetime}
\overline{Q}_{T}(\boldsymbol{\Omega},\vphi)=\sum_{\vk\in\ZZ^N}q_{\vk}(\boldsymbol{\Omega})
e^{i \vk\cdot\left(\vphi+\boldsymbol{\Omega}T/2 \right)}\mathrm{sinc}\left(\frac{\vk\cdot\boldsymbol{\Omega}T}{2}\right),
\end{equation}
where $\sinc(x)=\sin(x)/x$ denotes the unnormalized sinus cardinalis.
Since $\lim_{T\to \infty}\mathrm{sinc}\left(\alpha T/2\right)$ equals $1$
if $\alpha=0$ and $0$ otherwise, from equations~(\ref{ITA}) and
(\ref{finitetime}) we get 
\begin{equation}
\label{ITA1}
\overline{Q}_{\infty}(\boldsymbol{\Omega},\vphi)=\sum_{\vk\in\mathcal{S}_{\boldsymbol{\Omega}}}q_{\vk}(\boldsymbol{\Omega})e^{i \vk\cdot\vphi},
\end{equation}
where $\mathcal{S}_{\boldsymbol{\Omega}}$ is the set of all ordered $N$-tuples of integers orthogonal to $\boldsymbol{\Omega}$, i.e., $\mathcal{S}_{\boldsymbol{\Omega}}=\left\{\vk\in\ZZ^N: \vk\cdot\boldsymbol{\Omega}=0\right\}$.

If the $N$ components of the frequency vector $\boldsymbol{\Omega}$ are
incommensurable (i.e., it is not possible to express one of them as a
linear combination of the others with rational coefficients), then the set
$\mathcal{S}_{\boldsymbol{\Omega}}$ reduces to the single element
$\vk=\boldsymbol{0}$. In this case, from equation (\ref{ITA1}), it readily
follows that
\begin{equation}
\label{incommesurable}
\overline{Q}_{\infty}(\boldsymbol{\Omega},\vphi)=q_{\boldsymbol{0}}(\boldsymbol{\Omega})
\end{equation}
and, therefore, the infinite-time average is independent of $\vphi$. An
experimental verification of this general statement has been reported in
reference~\cite{Forster2015b}. 

In contrast, if the $N$ components of $\boldsymbol{\Omega}$ are
commensurable (i.e., it is possible to express one of them as a linear
combination of the others with rational coefficients), then the set
$\mathcal{S}_{\boldsymbol{\Omega}}$ contains additional elements other than
$\vk=\boldsymbol{0}$ and, according to equation (\ref{ITA1}),
$\overline{Q}_{\infty}(\boldsymbol{\Omega},\vphi)$ depends on the phases in
$\vphi$. In particular, if the frequencies are pairwise
commensurable, then there exists a frequency $\Omega$ such that
$\boldsymbol{\Omega}=\Omega \boldsymbol{n}$, where $\boldsymbol{n}=(n_1,\dots,n_N)$, with
$n_j$ being positive integers. Consequently, the condition
$\vk\cdot\boldsymbol{\Omega}=0$ becomes equivalent to the Diophantine
equation $\vk\cdot\boldsymbol{n}=0$. The general solution of the latter
equation can be expressed as an integer linear combination of a set of
$N-1$ generating vectors,
$\boldsymbol{g}^{(1)},\dots,\boldsymbol{g}^{(N-1)}$, each of which
satisfies the equation
$\boldsymbol{g}^{(j)}\cdot\boldsymbol{n}=0$~\cite{CasadoPascual2015a,
Morito1979a, Morito1980a}. Thus, equation (\ref{ITA1}) can be written as
\begin{equation}
\label{ITAmutcomm}
\overline{Q}_{\infty}(\boldsymbol{\Omega},\vphi)=\sum_{\boldsymbol{\ell}\in\ZZ^{N-1}}q_{\vk(\boldsymbol{\ell})}(\boldsymbol{\Omega})e^{i \vk(\boldsymbol{\ell})\cdot\vphi},
\end{equation}
where $\vk(\boldsymbol{\ell})=\sum_{j=1}^{N-1} \ell_j \boldsymbol{g}^{(j)}$.

Let $\mathcal{W}_{\mathrm{C}}$ and $\mathcal{W}_{\mathrm{I}}$ be the sets
of all $\boldsymbol{\Omega}$ whose components are commensurable and
incommensurable, respectively. From the above results, it can easily be
shown that the infinite-time average
$\overline{Q}_{\infty}(\boldsymbol{\Omega},\vphi)$ is discontinuous on the
set $\mathcal{W}_{\mathrm{C}}$. Indeed, since the set
$\mathcal{W}_{\mathrm{I}}$ is dense in $\RR^N$, for any
$\boldsymbol{\Omega}_{\mathrm{c}}\in \mathcal{W}_{\mathrm{C}}$ one can find
a sequence $\{\boldsymbol{\Omega}^{(n)}\}_{n\in\NN}\subset
\mathcal{W}_{\mathrm{I}}$ such that $\lim_{n\to
	\infty}\boldsymbol{\Omega}^{(n)}=\boldsymbol{\Omega}_{\mathrm{c}}$. If
$\overline{Q}_{\infty}(\boldsymbol{\Omega},\vphi)$ were continuous at
$\boldsymbol{\Omega}_{\mathrm{c}}$, then 
\begin{equation}
\overline{Q}_{\infty}(\boldsymbol{\Omega}_{\mathrm{c}},\vphi)=\lim_{n\to \infty}\overline{Q}_{\infty}(\boldsymbol{\Omega}^{(n)},\vphi)=\lim_{n\to \infty} q_{\boldsymbol{0}}(\boldsymbol{\Omega}^{(n)})
\end{equation}
and, as a result,
$\overline{Q}_{\infty}(\boldsymbol{\Omega}_{\mathrm{c}},\vphi)$ would be
independent of $\vphi$. Obviously, this contradicts the fact that, since
$\boldsymbol{\Omega}_{\mathrm{c}}\in \mathcal{W}_{\mathrm{C}}$,
$\overline{Q}_{\infty}(\boldsymbol{\Omega}_{\mathrm{c}},\vphi)$ depends on
$\vphi$.  Hence, we can conclude that
$\overline{Q}_{\infty}(\boldsymbol{\Omega},\vphi)$ is discontinuous at
$\boldsymbol{\Omega}_{\mathrm{c}}$ and, thus, on the set
$\mathcal{W}_{\mathrm{C}}$. Taking into account that
$\mathcal{W}_{\mathrm{C}}$ is also dense in $\RR^N$, this last result
implies that the infinite-time average
$\overline{Q}_{\infty}(\boldsymbol{\Omega},\vphi)$ is a highly
discontinuous function of the frequency vector $\boldsymbol{\Omega}$. In
this context, the following questions arise: (i) Can this discontinuity
actually be observed? (ii) How does it manifest itself in practice?

\section{Long-time asymptotic behavior of $\overline{Q}_{T}$} % of $Q$}
\label{sec:finitetime}

In real situations, the physical quantity $Q$ is known only in a finite
time-interval. Therefore, the infinite time-average
$\overline{Q}_{\infty}$ can be calculated only approximately by taking a
sufficiently large value of $T$. Suppose we are interested in
analyzing the dependence of $\overline{Q}_{T}$ on
$\boldsymbol{\Omega}$ near a fixed frequency vector
$\boldsymbol{\Omega}_0$. More precisely, we focus on values
of $\boldsymbol{\Omega}$ such that
$|\boldsymbol{\Omega}-\boldsymbol{\Omega}_0|$ is of the same order of
magnitude as $T^{-1}$, where
$|\boldsymbol{\Omega}-\boldsymbol{\Omega}_0|=[\sum_{j=1}^N(\Omega_j-\Omega_{0,j})^2]^{1/2}$.
Then, as $T$ increases, the size of the region of interest
becomes smaller in inverse proportion to $T$.

By defining the dimensionless frequency vector
$\delta\boldsymbol{\tilde{\omega}}=(\boldsymbol{\Omega}-\boldsymbol{\Omega}_0)T$,
the finite time-average can be brought to the form
\begin{eqnarray}
\overline{Q}_{T}(\boldsymbol{\Omega},\vphi)&=&\overline{Q}_{T}(\boldsymbol{\Omega}_0+ \delta \boldsymbol{\tilde{\omega}}/T,\vphi)\nonumber \\&=&\overline{Q}_{\mathrm{as}}(\boldsymbol{\Omega}_0,\delta \boldsymbol{\tilde{\omega}},\vphi)+R_{T}(\boldsymbol{\Omega}_0,\delta \boldsymbol{\tilde{\omega}},\vphi),
\label{asymp0}
\end{eqnarray}
where 
\begin{equation}
\label{asymp1}
\overline{Q}_{\mathrm{as}}(\boldsymbol{\Omega}_0,\delta \boldsymbol{\tilde{\omega}},\vphi)= \lim_{T\to\infty}\overline{Q}_{T}(\boldsymbol{\Omega}_0+ \delta \boldsymbol{\tilde{\omega}}/T,\vphi)
\end{equation}
is the leading-order of the asymptotic behavior of the function
$\overline{Q}_{T}(\boldsymbol{\Omega}_0+ \delta
\boldsymbol{\tilde{\omega}}/T,\vphi)$ for $T\to \infty$, while
$\delta \boldsymbol{\tilde{\omega}}$ is held constant.  For the rest
$R_{T}(\boldsymbol{\Omega}_0,\delta \boldsymbol{\tilde{\omega}},\vphi)
=\overline{Q}_{T}(\boldsymbol{\Omega}_0+ \delta \boldsymbol{\tilde{\omega}}/T,\vphi)
-\overline{Q}_{\mathrm{as}}(\boldsymbol{\Omega}_0,\delta
\boldsymbol{\tilde{\omega}},\vphi)$, it readily follows that
\begin{equation}
\label{RTlimit}
\lim_{T\to\infty}R_{T}(\boldsymbol{\Omega}_0,\delta \boldsymbol{\tilde{\omega}},\vphi)=0,
\end{equation}
and together with equation \eqref{ITA}, that
\begin{equation}
\overline{Q}_{\mathrm{as}}(\boldsymbol{\Omega}_0,\boldsymbol{0},\vphi)=\overline{Q}_{\infty}(\boldsymbol{\Omega}_0,\vphi).
\end{equation}

To obtain an expression for
$\overline{Q}_{\mathrm{as}}(\boldsymbol{\Omega}_0, \delta \boldsymbol{\tilde{\omega}},\vphi)$,
we set $\boldsymbol{\Omega}=\boldsymbol{\Omega}_0+\delta
\boldsymbol{\tilde{\omega}}/T$ in equation (\ref{finitetime}), and insert the
resulting expression into equation (\ref{asymp1}).
Then, using again $\lim_{T\to \infty}\mathrm{sinc}\left(\alpha T/2\right) =
\delta_{\alpha,0}$, one obtains
\begin{equation}
\label{asymp2}
\overline{Q}_{\mathrm{as}}(\boldsymbol{\Omega}_0,\delta \boldsymbol{\tilde{\omega}},\vphi)=\sum_{\vk\in\mathcal{S}_{\boldsymbol{\Omega}}}q_{\vk}(\boldsymbol{\Omega}_0)e^{i \vk\cdot\left(\vphi+\delta \boldsymbol{\tilde{\omega}}/2\right)}\mathrm{sinc}\left(\frac{\vk\cdot\delta \boldsymbol{\tilde{\omega}}}{2}\right).
\end{equation}
Using equation (\ref{ITA1}), one sees that equation (\ref{asymp2}) can be written in
the more compact form
\begin{equation}
\label{asymp3}
\overline{Q}_{\mathrm{as}}(\boldsymbol{\Omega}_0,\delta \boldsymbol{\tilde{\omega}},\vphi)=\int_{0}^{1}d\lambda \,\overline{Q}_{\infty}(\boldsymbol{\Omega}_0,\vphi+\lambda \delta \boldsymbol{\tilde \omega}).
\end{equation}
Therefore, in the limit $T\to \infty$, the asymptotic behavior of
$\overline{Q}_{T}$ in a neighborhood of $\boldsymbol{\Omega}_0$
of size proportional to $T^{-1}$ is completely determined by the
infinite-time limit $\overline{Q}_{\infty}$ at $\boldsymbol{\Omega}_0$.
Notice that, according to the above results, no matter how large
$T$, it is always possible to find a neighborhood of
$\boldsymbol{\Omega}_0$ within which $\overline{Q}_{T}$ is
continuous. Thus, the discontinuity mentioned in the previous section is a
mathematical idealization that, owing to the necessarily finite measurement
time, cannot be observed in reality.  

From equation (\ref{asymp3}) there immediately follows an interesting
conclusion. If $\boldsymbol{\Omega}_{0}\in \mathcal{W}_{\mathrm{I}}$ (i.e.,
all $N$ components are incommensurable), then
$\overline{Q}_{\infty}(\boldsymbol{\Omega}_0,\vphi)$ is independent of
$\vphi$ and $\overline{Q}_{\mathrm{as}}(\boldsymbol{\Omega}_0,\delta
\boldsymbol{\tilde{\omega}},\vphi)= \overline{Q}_{\infty}(\boldsymbol{\Omega}_0,\vphi)$
for all $\delta \boldsymbol{\tilde{\omega}}$. Consequently, from equations~(\ref{ITA})
and (\ref{asymp1}), one obtains that
\begin{equation}
\lim_{T\to\infty}\left[\overline{Q}_{T}(\boldsymbol{\Omega}_0+ \delta \boldsymbol{\tilde{\omega}}/T,\vphi)-\overline{Q}_{T}(\boldsymbol{\Omega}_0,\vphi)\right]=0
\end{equation}
for all $\delta \boldsymbol{\tilde{\omega}}$. In contrast, if $\boldsymbol{\Omega}_{0}\in \mathcal{W}_{\mathrm{C}}$ (i.e., its $N$ components are commensurable), then $\overline{Q}_{\infty}(\boldsymbol{\Omega}_0,\vphi)$ depends on $\vphi$ and, in general, 
\begin{equation}
\lim_{T\to\infty}\left[\overline{Q}_{T}(\boldsymbol{\Omega}_0+ \delta \boldsymbol{\tilde{\omega}}/T,\vphi)-\overline{Q}_{T}(\boldsymbol{\Omega}_0,\vphi)\right]\neq 0.
\end{equation}
This difference in behavior will be apparent in the numerical calculations
presented below.

\section{Bichromatic driving}
\label{sec:bichromatic}

The analytic considerations made so far provide the long-time limit of the
non-linear response to a multi-frequency forcing.  In order to investigate
numerically how this limit is approached, we focus on bichromatic driving,
i.e., on dynamic equations which contain terms of the form $f_1(t)=A_1
\cos(\Omega_1 t+\varphi_1)$ and  $f_2(t)=A_2 \cos(\Omega_2 t+\varphi_2)$,
where our line of reasoning still holds if the cosines are replaced by any
other $2\pi$-periodic functions.

\subsection{Commensurable frequencies and periodicity in $\boldsymbol{\varphi}$}

\label{ComFreq}

While the $2\pi$-periodicity of $\{f_1(t),f_2(t)\}$ in $\varphi_1$ and $\varphi_2$
is evident, discrete time translation symmetry is present only when $\Omega_1$ and $\Omega_2$ are commensurable, i.e., for rational values of 
$\Omega_2/\Omega_1$. Indeed, let us assume that $f_1(t)$ and $f_2(t)$ have
a common fundamental period $\mathcal{T}$. From the periodicity of the
cosine function, it then follows that there must exist two integers $q$ and
$p$ such that $\Omega_1 \mathcal{T}=2\pi q$ and $\Omega_2\mathcal{T}=2\pi
p$. This implies that $\Omega_1 =q\Omega$ and $\Omega_2=p \Omega$, with
$\Omega=2\pi/\mathcal{T}$, and consequently, that $\Omega_2/\Omega_1=p/q$.
In addition, the integers $q$ and $p$ must be coprime because otherwise
$\mathcal{T}/\gcd(q,p) < \mathcal{T}$, would be the common fundamental
period of $f_1(t)$ and $f_2(t)$, with $\gcd(q,p)$ denoting the greatest
common divisor of $q$ and $p$.

Let us now focus on a frequency vector $\boldsymbol{\Omega}_0\equiv (q ,p)
\Omega$, with some coprime integers $p$ and $q$, henceforth referred to as
$(q,p)$-resonance. Using equation (\ref{Fourier1}), it can easily be seen
that the stationary value
$Q_t^{{{\mathrm{st}}}}(\boldsymbol{\Omega}_0,\vphi)$ is a
$\mathcal{T}$-periodic function of time. A further interesting fact is that
the infinite-time average
$\overline{Q}_{\infty}(\boldsymbol{\Omega}_0,\vphi)$ displays a higher
symmetry in $\vphi$ than the obvious $\varphi_1 \to \varphi_1+2\pi$ and
$\varphi_2 \to \varphi_2+2\pi$~\cite{CasadoPascual2013a}. To see that this
is so, note that in the present case the condition
$\vk\cdot\boldsymbol{\Omega}_0=0$ becomes equivalent to $k_1 q +k_2 p=0$.
The general solution of this Diophantine equation is $\ell \boldsymbol{g}$,
with the generating vector $\boldsymbol{g}\equiv(-p,q)$ and any integer
$\ell$. Thus, in this case, equation (\ref{ITAmutcomm}) becomes
\begin{equation}
\label{2freqcomm}
\overline{Q}_{\infty}(\boldsymbol{\Omega}_0,\vphi)=\sum_{\ell\in\ZZ}q_{\ell \boldsymbol{g}}(\boldsymbol{\Omega}_0)e^{i \ell  \boldsymbol{g}\cdot \vphi}.
\end{equation}
Since $\boldsymbol{g}\cdot \vphi=-p\varphi_1+q\varphi_2$, from equation (\ref{2freqcomm}) it readily follows that $\overline{Q}_{\infty}(\boldsymbol{\Omega}_0,\vphi)$ is $2\pi/p$-periodic in $\varphi_1$ and $2\pi/q$-periodic in $\varphi_2$.

\subsection{The neighborhood of commensurable frequencies}

Let us now turn our attention to the vicinity of a $(q,p)$-resonance. More
specifically, and following the general approach outlined in
Section~\ref{sec:finitetime}, we focus on values of $\boldsymbol{\Omega}$
such that $|\boldsymbol{\Omega}-\boldsymbol{\Omega}_0|=[(\Omega_1-q
\Omega)^2+(\Omega_2-p \Omega)^2]^{1/2}$ is of the same order as $T^{-1}$.
Then, introducing the notation
$\delta\boldsymbol{\omega}=\delta\boldsymbol{\tilde{\omega}}/T=\boldsymbol{\Omega}-\boldsymbol{\Omega}_0$,
from equation ~(\ref{asymp0}), (\ref{RTlimit}), and (\ref{asymp2}), it follows that
\begin{multline}
\label{asympbic}
\overline{Q}_{T}(\boldsymbol{\Omega}_0+\delta\boldsymbol{\omega},\vphi)\thicksim\sum_{\ell\in\ZZ}q_{\ell \boldsymbol{g}}\left(\boldsymbol{\Omega}_0\right)e^{i \ell \boldsymbol{g}\cdot\left(\vphi+T\delta\boldsymbol{\omega}/2\right)}
\\ \times \mathrm{sinc}\left(\frac{\ell T \boldsymbol{g} \cdot \delta\boldsymbol{\omega}}{2}\right),
\end{multline}
provided that $T$ is large enough so that the term $R_{T}$ can be neglected.

Two conclusions can be drawn from equation (\ref{asympbic}). First, the time
average
$\overline{Q}_{T}(\boldsymbol{\Omega}_0+\delta\boldsymbol{\omega},\vphi)$
is $2\pi/p$-periodic in $\varphi_1$ and $2\pi/q$-periodic in $\varphi_2$.
There is, however, an important difference to the exact periodicity of
the infinite-time average
$\overline{Q}_{\infty}(\boldsymbol{\Omega}_0,\vphi)$ presented in
Section~\ref{ComFreq}. Here the periodicity holds only for sufficiently large
values of $T$ and for $\boldsymbol{\Omega}$ sufficiently close to
$\boldsymbol{\Omega}_0$---to be precise, for
$|\delta\boldsymbol{\omega}|=\mathcal{O}(T^{-1})$. Second, under these same
restrictions, the relative height $\Delta
\overline{Q}_{T}(\boldsymbol{\Omega}_0+\delta\boldsymbol{\omega},\vphi)\equiv
\overline{Q}_{T}(\boldsymbol{\Omega}_0+\delta\boldsymbol{\omega},\vphi)-q_{\boldsymbol{0}}\left(\boldsymbol{\Omega}_0\right)$
vanishes when all $\sinc$ functions in equation \eqref{asympbic} are zero
for all $\ell\neq 0$, i.e., when
$\boldsymbol{g}\cdot\delta\boldsymbol{\omega}$ is a nonzero integer
multiple of $2\pi/T$. In particular, if we set $\delta \omega_1=0$ and vary
$\delta \omega_2$, $\Delta
\overline{Q}_{T}(\boldsymbol{\Omega}_0+\delta\boldsymbol{\omega},\vphi)$
vanishes when $\delta \omega_2$ is a nonzero integer multiple of $2\pi/(q
T)$. Analogously,  if we set $\delta \omega_2=0$ and vary $\delta
\omega_1$, $\Delta
\overline{Q}_{T}(\boldsymbol{\Omega}_0+\delta\boldsymbol{\omega},\vphi)$
vanishes when $\delta \omega_1$ is a nonzero integer multiple of $2\pi/(p
T)$. Hence,  in the vicinity of the $(q,p)$-resonance, $2\pi/(q T)$ and
$2\pi/(p T)$ represent the frequency scales on which $\Delta
\overline{Q}_{T}(\boldsymbol{\Omega}_0+\delta\boldsymbol{\omega},\vphi)$
varies.

The main aim of the next two sections is to provide quantitative
information on how these asymptotic properties can actually be observed in
practice. Note, for example, that a further limitation may stem from the
fact that the set of commensurable frequencies vectors is dense in $\RR^2$
and, consequently, any $(q,p)$-resonance may be disturbed by another
resonance that lies arbitrarily close to it. It will turn out, however, for
all cases investigated, practically only resonances with rather small $q$
and $p$ matter. Moreover, for sufficiently large values of $T$,  the shape
of the resonance peak is mainly governed by the term with $\ell=1$ in
equation \eqref{asympbic} and, therefore, their enveloping function appears as a
rather clean $\sinc$ function.

\section{Classical random walk model}
\label{sec:randomwalk}

In the model presented in this section, the motion of a particle
in a periodic substrate is given by a random walk on a one-dimensional
lattice. The lattice sites are located at $x_n=n a$, where $n$ is any integer 
and $a$ the distance between two neighboring sites.  The evolution of the
probabilities $p_n(t)$ that the particle is at site $n$ is governed by the
master equation~\cite{vanKampen1992a}
\begin{eqnarray}
\label{Mastereq}
\dot{p}_n(t)&=&-\left[r_{+}(t)+r_{-}(t)\right]p_n(t)\nonumber\\
&&+r_{+}(t)p_{n-1}(t)+r_{-}(t)p_{n+1}(t),
\end{eqnarray}
where $r_{+}(t)$ and $r_{-}(t)$ are the transition rates from site $n$ to
site $n+1$ and $n-1$, respectively.  They are assumed to be independent of
$n$ and to follow the Van't Hoff-Arrhenius law~\cite{Hanggi1990a}
\begin{equation}
\label{Arrhenius}
r_{\pm}(t)=r_0 e^{-\beta \left[E_0 \pm \Delta E(t)\right]},
\end{equation}
where $r_0$ is a prefactor with the dimension of an inverse time, $\beta=(k_B
\Theta)^{-1}$ is the inverse temperature, and $E_0 + \Delta E(t)$ and $E_0
- \Delta E(t)$ are, respectively, the activation energies for the forward
and backward steps. These activation energies oscillate around a constant
value $E_0$ with time-dependent amplitudes $\Delta E(t)$ and $-\Delta
E(t)$, respectively.

In this section, we will consider that the role of $Q_t$ is played by  the
mean particle velocity $V_t$, which is defined as the time derivative of
the mean particle position $X_t=a \sum_{n\in\ZZ}n p_n(t)$. Using
equations \eqref{Mastereq} and \eqref{Arrhenius}, it can be seen easily that
\begin{equation}
\label{Vt}
V_t=a \left[r_{+}(t)-r_{-}(t)\right]=v \sinh\left[f(t)\right],
\end{equation}
where $v=2 a r_0 e^{-\beta E_0}$ and $f(t)=-\beta \Delta E(t)$. Henceforth, we will assume that 
\begin{equation}
\label{deff}
f(t)=A_1 \cos (\Omega_1 t+\varphi_1)+A_2 \cos (\Omega_2 t+\varphi_2),
\end{equation}
with $A_1$ and $A_2$ being two dimensionless constants, which can be taken
as positive by suitable choice of the phases $\varphi_1$ and $\varphi_2$.
Note that, in the present model, there is no difference between the
stationary $V_t^{\mathrm{st}}$ and $V_t$ because the mean particle velocity
is independent of the initial preparation. In addition, from
equations \eqref{Vt} and \eqref{deff}, it immediately follows that $V_t$
satisfies the symmetry properties
\begin{align}
\label{sym1}
V_{-t}(\boldsymbol{\Omega},-\vphi)={}& V_{t}(\boldsymbol{\Omega},\vphi) ,
\\
\label{sym2}
V_{t}(\boldsymbol{\Omega},\vphi+\boldsymbol{\pi})={}& -V_{t}(\boldsymbol{\Omega},\vphi),
\end{align}
where $\boldsymbol{\pi}\equiv(\pi,\pi)$.

Now the Fourier expansion in equation (\ref{Fourier1}) takes the form
\begin{equation}
\label{RWresult0}
V_{t}(\boldsymbol{\Omega},\vphi)=\sum_{\vk\in \ZZ^2}v_{\vk}\,e^{i \vk\cdot\left(\vphi+\boldsymbol{\Omega}t \right)},
\end{equation}
with
\begin{equation}
\label{vk1}
v_{\vk}=v \int_{-\pi}^{\pi}\int_{-\pi}^{\pi} \frac{e^{-i\vk\cdot\vphi} }{(2\pi)^2 }\,\sinh\left(A_1 \cos \varphi_1+A_2 \cos \varphi_2\right) d^2\vphi.
\end{equation}
Since the Fourier expansion is unique, using equations (\ref{sym1}) and
(\ref{RWresult0}), it is easy to see that $v_{-\vk}=v_{\vk}$. For the same
reason, from equations (\ref{sym2}) and (\ref{RWresult0}), it readily
follows that $v_{\vk}=-e^{i(k_1+k_2)\pi}v_{\vk}$. In addition, from
equation (\ref{vk1}),  it is clear that $v_{\vk}^{\ast}=v_{-\vk}$, where the asterisk denotes complex conjugation. In conclusion, all the coefficients $v_{\vk}$ are real and vanish when $k_1+k_2$ is an even integer.

With the above results, let us now examine the dependence of the
infinite-time average velocity on $\boldsymbol{\Omega}$. If $\Omega_1$ and
$\Omega_2$ are incommensurable, from equation (\ref{incommesurable}) one concludes that $\overline{V}_{\infty}(\boldsymbol{\Omega},\vphi)=v_{\boldsymbol{0}}=0$. If, by contrast,  $\Omega_1$ and $\Omega_2$ are commensurable, taking into account that $v_{\vk}=v_{-\vk}$ and that $v_{2\vk}=0$, 
it is easy to see from equation (\ref{2freqcomm}) that
\begin{equation}
\label{Vinftyrw}
\overline{V}_{\infty}(\boldsymbol{\Omega}_0,\vphi)=2\sum_{\ell=0}^{\infty}v_{(2\ell+1)\boldsymbol{g}}\cos\left[(2\ell+1) \boldsymbol{g} \cdot \vphi \right].
\end{equation}
Using the definition of $\boldsymbol{g}$ and the fact that $v_{\vk}$
vanishes when $k_1+k_2$ is even, it can be verified with
equation (\ref{Vinftyrw}) that
$\overline{V}_{\infty}(\boldsymbol{\Omega}_0,\vphi)=0$  when $q-p$ is even.
Therefore, as a consequence of the symmetry property~(\ref{sym2}), only the
$(q,p)$-resonances with $q-p$ odd are present in this model.

The integral in equation (\ref{vk1}) can be evaluated explicitly by expanding
the hyperbolic sine into a power series. Then, after some calculations, we
obtain
\begin{multline}
\label{vk}
v_{\vk}=v\sum_{j=0}^{\infty}\sum_{\ell=0}^{2j+1}\sum_{m=0}^{\ell}\sum_{n=0}^{2j+1-\ell}\delta_{k_1,2m-\ell}\delta_{k_2,2n+\ell-2j-1}\\
\times \frac{A_1^{\ell} A_2^{2j+1-\ell}}{2^{2j+1}m!n!(\ell-m)!(2j+1-\ell-n)!}.
\end{multline}

It can be verified that the coefficients $v_{\vk}$ given by equation (\ref{vk}),
as could not be otherwise, satisfy the conditions discussed above. In
addition, since $A_1$ and $A_2$ are positive, all the coefficients
$v_{\vk}$ are clearly non-negative. Thus, from equation (\ref{Vinftyrw}), it
follows that the maximum value of $\overline{V}_{\infty}(\boldsymbol{\Omega}_0,\vphi)$ is 
\begin{equation}
\label{MaxHeight}
\overline{V}_{\infty,\mathrm{M}}(\boldsymbol{\Omega}_0)=2\sum_{\ell=0}^{\infty}v_{(2\ell+1)\boldsymbol{g}},
\end{equation}
and it occurs when $\boldsymbol{g}\cdot\vphi$ is an integer multiple of $2\pi$, i.e., when 
\begin{equation}
\label{maxres}
q\varphi_2-p\varphi_1=2\pi n
\end{equation}
with $n$ being any integer.  Assuming that $\varphi_1$ and $\varphi_2$
satisfy this condition of maximum resonance, from equation (\ref{asympbic})
together with $v_{-\vk}=v_{\vk}$, it follows that
\begin{equation}
\label{asympbic2}
\overline{V}_{T}(\boldsymbol{\Omega}_0+\delta\boldsymbol{\omega},\vphi)\thicksim 2\sum_{\ell=0}^{\infty}v_{(2\ell+1)\boldsymbol{g}} \,\mathrm{sinc}\left[(2\ell+1) T\boldsymbol{g}\cdot \delta\boldsymbol{\omega}\right],
\end{equation}
provided that $|\delta\boldsymbol{\omega}|=\mathcal{O}(T^{-1})$ and that
$T$ is large enough such that the term $R_{T}$ can be neglected. Note that,
in this case,
$\overline{V}_{T}(\boldsymbol{\Omega}_0+\delta\boldsymbol{\omega},\vphi)$
vanishes whenever $\boldsymbol{g}\cdot \delta\boldsymbol{\omega}$ is a
nonzero integer multiple of $\pi/T$. Thus, if we set $\delta \omega_1=0$
and vary $\delta \omega_2$,
$\overline{V}_{T}(\boldsymbol{\Omega}_0+\delta\boldsymbol{\omega},\vphi)$
vanishes when $\delta \omega_2$ is a nonzero integer multiple of $\pi/(q
T)$, whereas if we set $\delta \omega_2=0$ and vary $\delta \omega_1$,
$\overline{V}_{T}(\boldsymbol{\Omega}_0+\delta\boldsymbol{\omega},\vphi)$
vanishes if $\delta \omega_1$ is a nonzero integer multiple of $\pi/(p T)$.

To illustrate our theoretical results, we have calculated
$\overline{V}_{T}(\boldsymbol{\Omega},\vphi)$ using equation
(\ref{finitetime}) with
$\overline{Q}_{T}(\boldsymbol{\Omega},\vphi)=\overline{V}_{T}(\boldsymbol{\Omega},\vphi)$,
and $q_{\vk}(\boldsymbol{\Omega})=v_{\vk}$.  The coefficients $v_{\vk}$
appearing in that equation have been evaluated using equation (\ref{vk}). To
ensure that both drivings have roughly the same influence, we have focused
on the most symmetric case of equal driving amplitudes.  Specifically, in
all the figures of this section we have taken $A_1=A_2=1$. In addition, to
maximize the height of the resonance peaks, we have restricted our analysis
to values of $\varphi_1$ and $\varphi_2$ that satisfy the condition of
maximum resonance in equation (\ref{maxres}).

\begin{figure}
	\centerline{\includegraphics[width=87mm]{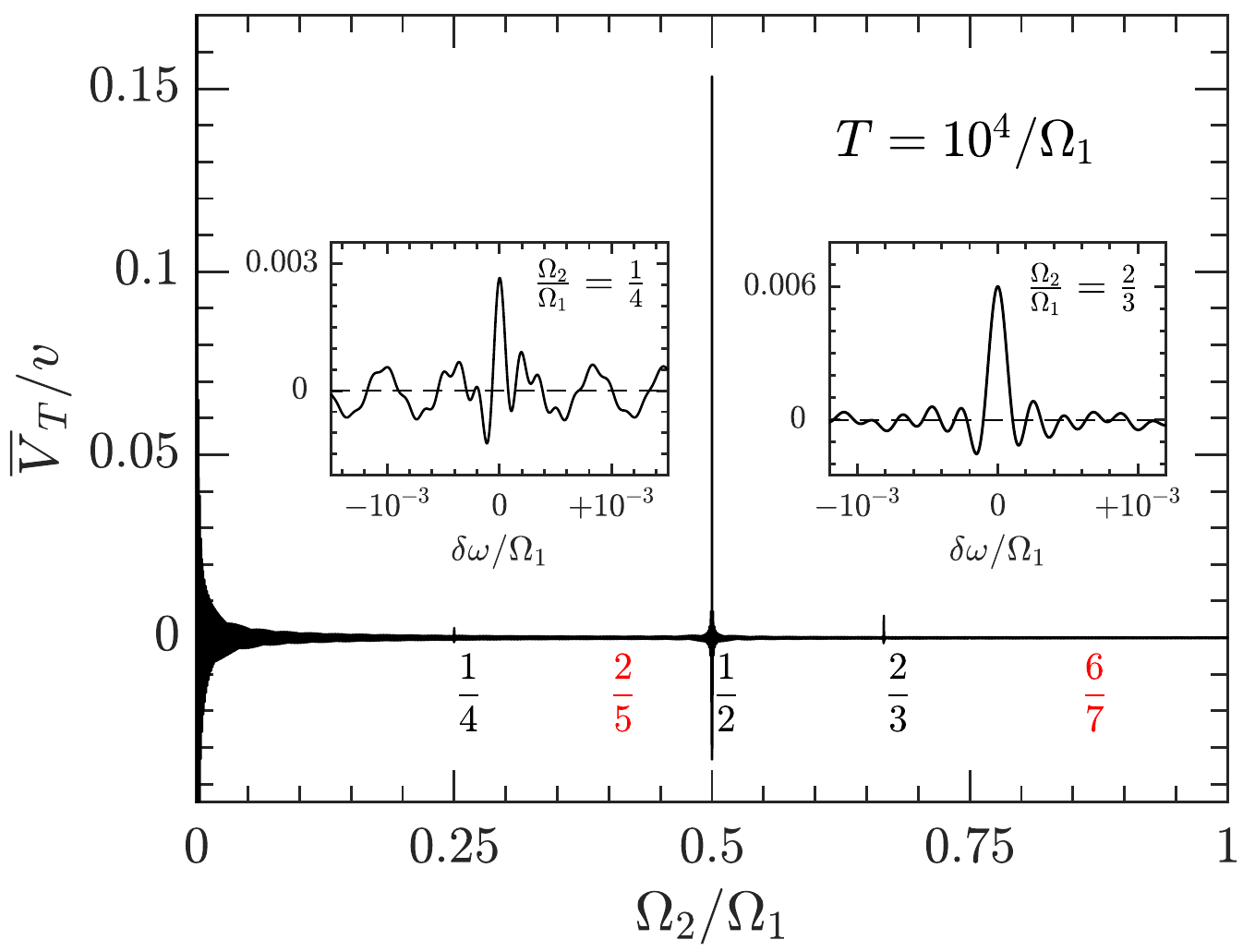}}
	\caption{Dependence of the dimensionless average velocity $\overline{V}_{T}/v$ on $\Omega_2/\Omega_1$ for $A_1=A_2=1$, $\varphi_1=\varphi_2=0$, and $T=10^4/\Omega_1$. The resonance peaks corresponding to the fractions $0$ (only partially shown for better visibility of the remaining peaks) and $1/2$ are clearly visible. The ones corresponding to $1/4$ and $2/3$ are also visible but considerably smaller. Other resonances, such as the ones corresponding to $2/5$ and $6/7$ shown in Figure~\ref{fig:otra2} cannot be appreciated on this scale. The insets show a zoomed-in view around $\Omega_2/\Omega_1=1/4$ (left inset) and $\Omega_2/\Omega_1=2/3$ (right inset) in terms of $\delta \omega/\Omega_1 =\Omega_2/\Omega_1-p/q$.}
	\label{fig:barrido}
\end{figure}

In Figure~\ref{fig:barrido}, we plot the dimensionless time-average velocity
$\overline{V}_{T}/v$ as a function of $\Omega_2/\Omega_1$ for
$\varphi_1=\varphi_2=0$ and $T=10^4/\Omega_1$. According to our theoretical
results, there should emerge resonance peaks when $\Omega_2/\Omega_1$ is
equal to a rational number. Furthemore, only the resonances corresponding
to the fractions $0$, $1/4$, $1/2$, and $2/3$ are visible. Other
resonances, such as the ones corresponding to $2/5$ and $6/7$ studied below, cannot be
appreciated in the figure. This absence of resonances is not due to the use
of a finite averaging time. In fact, these peaks would be imperceptible
even in the limit $T\to\infty$, because they are very small compared to the
smallest peak visible in Figure~\ref{fig:barrido}. Indeed, using
equation (\ref{MaxHeight}), it is easy to verify that at $\Omega_2/\Omega_1=1/4$
(the smallest peak visible in Figure~\ref{fig:barrido}),
$\overline{V}_{\infty,\mathrm{M}}$ is approximately equal to $3\times
10^{-3}v$, whereas at $\Omega_2/\Omega_1=2/5$ and  $\Omega_2/\Omega_1=6/7$,
$\overline{V}_{\infty,\mathrm{M}}$ is approximately equal to $7\times
10^{-5}v$ and $7\times 10^{-11}v$, respectively.

To analyze in more detail the behavior of $\overline{V}_{T}$ in the
vicinity of a $(q,p)$-resonance, we now consider that one frequency,
say $\Omega_1$, is kept fixed, while the other frequency, $\Omega_2$, is
varied around the value $p \Omega_1/q$. In terms of our previous notation,
this corresponds to setting $\delta \omega_1=0$ and $\delta
\omega_2=\Omega_2-p \Omega_1/q$. Henceforth, for notational simplicity, we
will write $\delta \omega$ instead of $\delta \omega_2$. In addition, to
facilitate comparison with the asymptotic expression~(\ref{asympbic2}), we
will use the dimensionless variable $q \delta \omega T$, which here is nothing
but $T\boldsymbol{g}\cdot \delta\boldsymbol{\omega}$. 

\begin{figure}
	\centerline{\includegraphics[width=87mm]{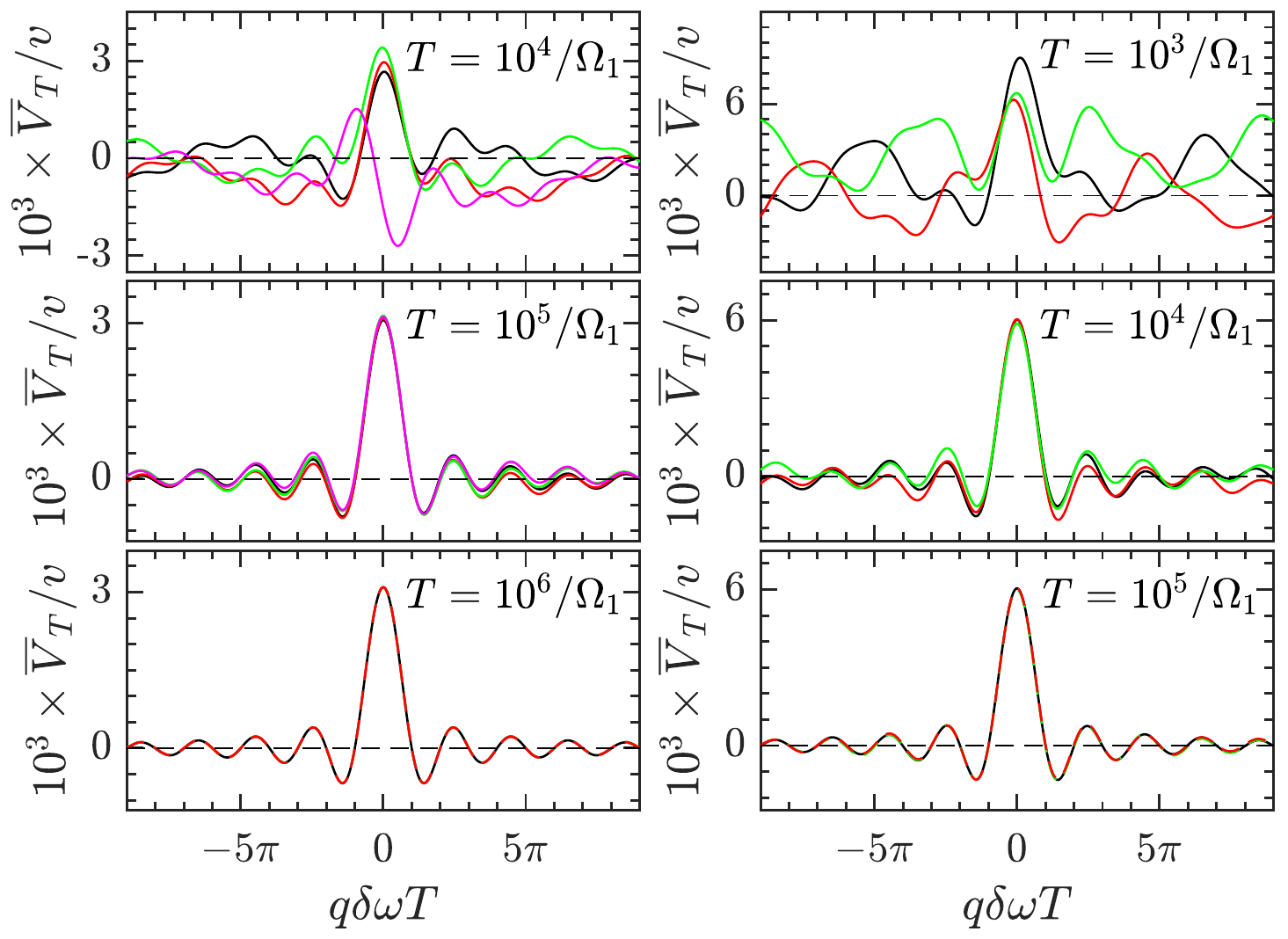}}
	\caption{Dependence of the dimensionless average velocity $\overline{V}_{T}/v$ on the dimensionless variable $q \delta \omega T$ for the resonances $(q,p)=(4,1)$ (left column) and $(q,p)=(3,2)$ (right column), and the values of the averaging times displayed
		in the panels. For all the curves $\varphi_1=0$ and
$A_1=A_2=1$. In each panel, there are $q$ curves corresponding to the
values $\varphi_2=2\pi n/q$, for $n=0,\dots,q-1$, which are obtained from
the condition of maximum resonance in equation (\ref{maxres}). As expected
from the theoretical analysis, with increasing the value of $T$, the $q$
curves converge to the asymptotic curve given by equation (\ref{asympbic2}).}
	\label{fig:otra1}
\end{figure}

\begin{figure}
	\centerline{\includegraphics[width=87mm]{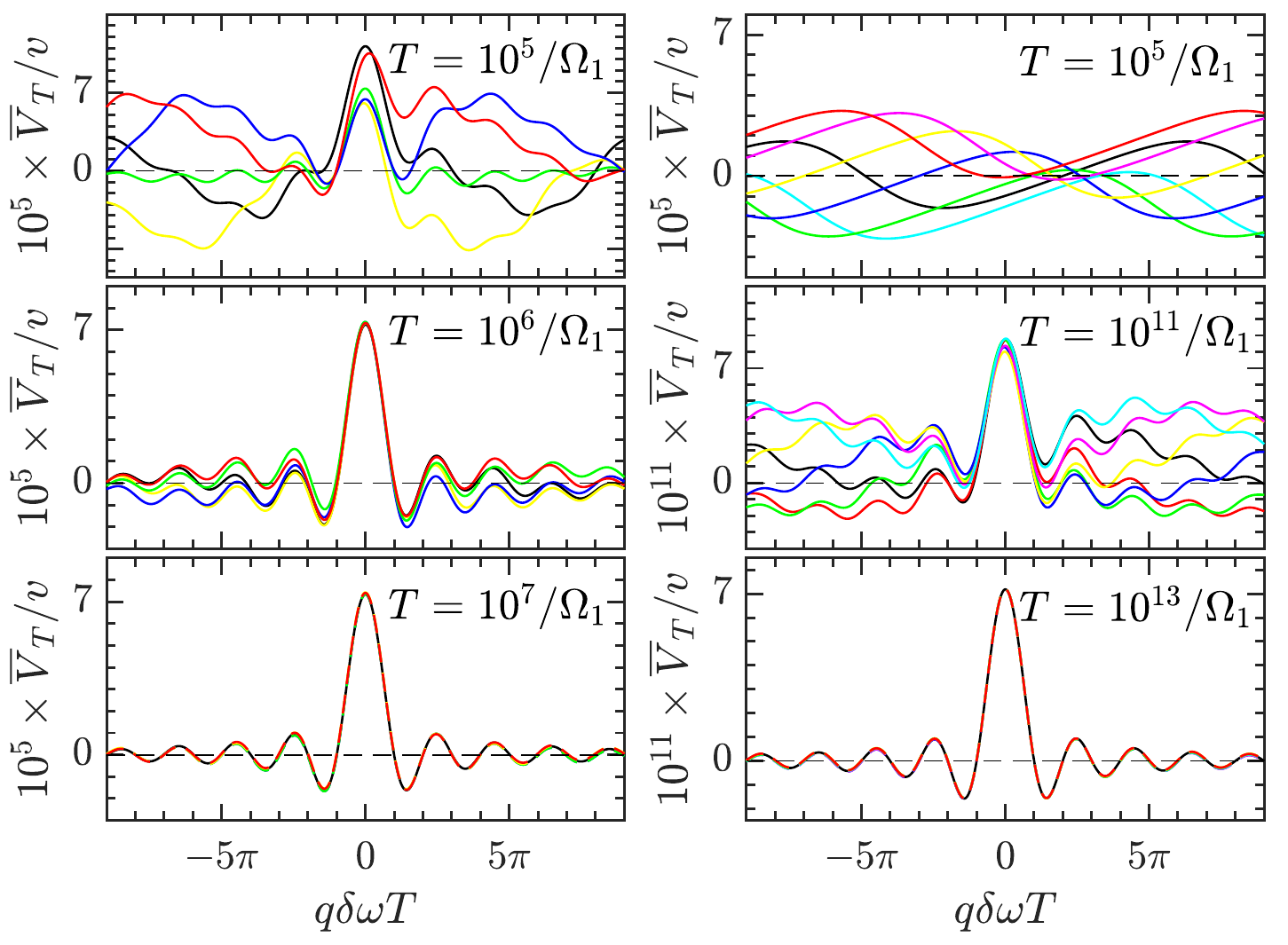}}
	\caption{The same as in Figure~\ref{fig:otra1} but for the resonances $(q,p)=(5,2)$ (left column) and $(q,p)=(7,6)$ (right column).}
	\label{fig:otra2}
\end{figure}

In Figures~\ref{fig:otra1} and \ref{fig:otra2}, we depict the dependence of
$\overline{V}_{T}/v$ on $q \delta \omega T$ for the resonances
$(q,p)=(4,1)$ (left column in Figure~\ref{fig:otra1}), $(q,p)=(3,2)$ (right
column in Figure~\ref{fig:otra1}), $(q,p)=(5,2)$ (left column in
Figure~\ref{fig:otra2}), and $(q,p)=(7,6)$ (right column in
Figure~\ref{fig:otra2}). To analyze how these peaks emerge as the averaging
time increases, different values of $T$ have been considered, which are
indicated in the panels. In addition, in each panel, $q$ curves have been
plotted, corresponding to the values $\varphi_2=2\pi n/q$, with
$n=0,\dots,q-1$. Since $\varphi_1=0$, all these values satisfy the
condition of maximum resonance in equation (\ref{maxres}). The results in
Figures~\ref{fig:otra1} and \ref{fig:otra2} corroborate the theoretical
prediction that, for sufficiently large values of $T$, the $q$ curves
converge to the asymptotic result in equation (\ref{asympbic2}). Surprisingly,
the averaging times necessary to reach the asymptotic regime are huge in
comparison to the period of the driving $\mathcal{T}=2\pi q/\Omega_1$.   

\section{Quantum mechanical two-level system}
\label{sec:tls}

Let us now consider a dissipative quantum mechanical model with the
Hamiltonian
$H(t) = H_0 + H_D(t)$, where
\begin{equation}
H_0 = \frac{\epsilon}{2}(\sigma_x \cos\theta + \sigma_z \sin\theta)
\end{equation}
and the bichromatic driving
\begin{align}
H_D(t) ={}& A_1\sigma_x \cos(\Omega_1 t +\varphi_1) + A_2\sigma_z\cos(\Omega_2 t+\varphi_2) .
\end{align}
To ensure that both drivings have roughly the same impact, we focus on the
most symmetric case $\theta = \pi/4$ and equal driving amplitudes,
$A_1=A_2\equiv A$.

For the consideration of dissipation, one may start from a system-bath
model to obtain an equation of motion for the reduced density operator of
the dissipative system.  Then one can show that generally dissipation is
quantitatively affected by the driving \cite{Kohler1997a, Grifoni1998a}.
Here however, we are interested in the generic response to bichromatic
drivings and, thus, we follow a less involved path which allows an efficient
numerical solution for rather long propagation times.  In doing so, we 
employ a Lindblad master equation for the
density operator \cite{Breuer2003a}, $\dot\rho = -i[H_0+H_D(t), \rho]
+\gamma\mathcal{D}(\rho)$ (in units with $\hbar=1$) with a dissipator
\cite{Breuer2003a}
\begin{equation}
\mathcal{D} \rho = \tilde\sigma_-\rho \tilde\sigma_+
-\frac{1}{2}\tilde\sigma_+ \tilde\sigma_-\rho
-\frac{1}{2}\rho \tilde\sigma_+ \tilde\sigma_- \,,
\end{equation}
where $\tilde\sigma_- = |\varphi_0\rangle\langle\varphi_1| =
\sigma_+^\dagger$ induces dissipative decay from the excited state
$|\varphi_1\rangle$ of the undriven Hamiltonian $H_0$ to the corresponding
ground state $|\varphi_0\rangle$.

As an observable $Q$, we may choose any combination of Pauli matrices.
Generic features, however, will not depend on the particular choice such that
without loss of generality, we consider
$Q_t^\mathrm{st}\equiv\langle\sigma_z\rangle_t$.  To ensure independence of details,
we have verified all numerical results by using also slightly different
setups and observables.

\begin{figure}
\centerline{\includegraphics{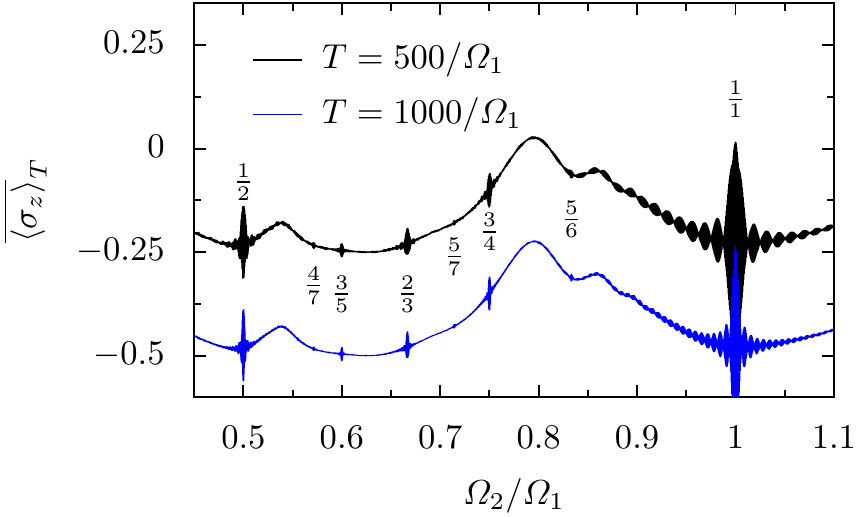}}
\caption{Time-averaged response of the bichromatically driven two-level
system after a transient stage averaged over various times $T$.  When
$\Omega_2/\Omega_1$ is close to a rational $p/q$, peaks with generic shape
emerge (labeled by the corresponding fraction).
Parameter values are: $\epsilon=\sqrt{2}\Omega_1$, $\theta=\pi/4$,
$A_1=A_2=\Omega_1$, and $\gamma=0.2\Omega_1$.
For graphical reasons, the curve for $T=500/\Omega_1$ is vertically shifted.}
\label{fig:overview}
\end{figure}

We start by sketching the global picture of the response
(Figure~\ref{fig:overview}) which shows
$\overline{\langle\sigma_z\rangle}_T$ for two different averaging times $T$
as a function of $\Omega_2$ and for various phases $\varphi_2$ (in this
section, we always take $\varphi_1=0$).  As expected,
$\overline{\langle\sigma_z\rangle}_T$ exhibits resonance peaks at simple
rational values of $\Omega_2/\Omega_1$ which sharpen with increasing $T$.
The size of the peaks as well as the shape of the background depend on
details.  Generally it is such that the (1,1) resonance is rather
prominent, which can be understood by frequency mixing: For two equal
driving frequencies, a zero-frequency response can be obtained already in
second-order perturbation theory.  For all other values of $(q,p)$, one has
to go to higher order.  Henceforth we focus on the universal features in
narrow regions around pronounced peaks.

\begin{figure}
\centerline{\includegraphics{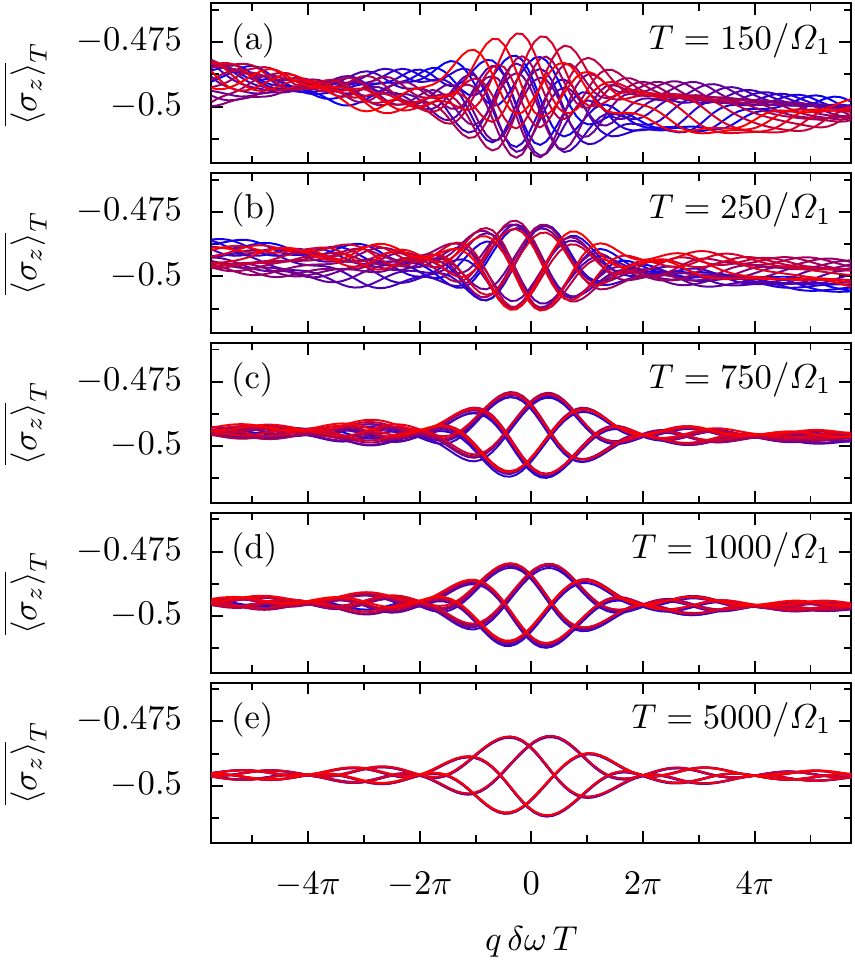}}
\caption{Resonance peak $(q,p) = (5,3)$ for the phases $\varphi_1=0$ and
$\varphi_2=2\pi n/(4q)$, with $n=0,1,..,4q-1$, showing the transition from
$2\pi$-periodicity to $2\pi/q$-periodicity.  Notice that the abscissa is
scaled with the averaging time $T$.  All other parameters are as in
Figure~\ref{fig:overview}.}
\label{fig:buildupall}
\end{figure}

As smaller peaks tend to be less compromised by components with $\ell\neq
1$, we study the emergence of a peak with increasing propagation time $T$
for $(q,p) = (5,3)$.   Figure~\ref{fig:buildupall} shows
$\overline{\langle\sigma_z\rangle}_T$
for the 20 equally spaced phases $\varphi_2 = 2\pi n/20$ with
$n=0,1,\ldots, 19$.  For the relatively short propagation time
$T=150/\Omega_1$, all curves are significantly different from each other,
while a clear peak structure is still missing.  With increasing $T$ and
staring at the center $\delta\omega=0$, curves for phases that differ by
$2\pi/q$ coincide, such that eventually $20/q = 4$ groups of curves emerge.
This reflects the $2\pi/q$ periodicity in $\varphi_2$ derived above for
$\delta\omega=0$.  Moreover, it confirms the generalization conjectured
from equation \eqref{asympbic}, namely that the periodicity to a good
approximation holds in a whole neighborhood of the $(q,p)$-peak.

\begin{figure}
\centerline{\includegraphics{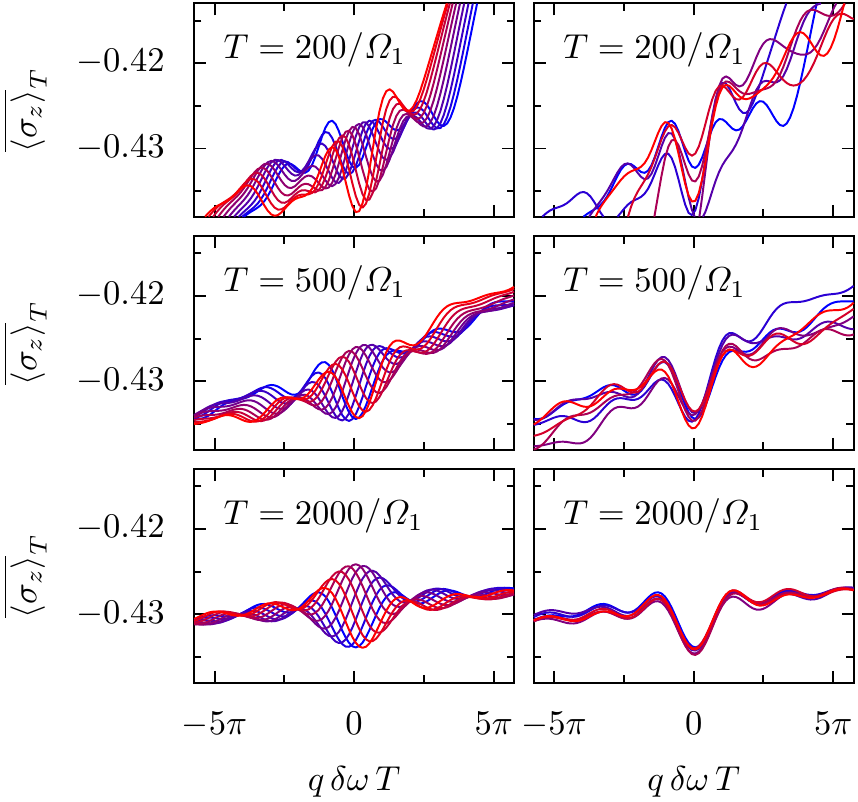}}
\caption{Resonance peak $(q,p)=(7,5)$ for the averaging times displayed
in the graphics, while all other parameters are as in
Figure~\ref{fig:overview}.
Left column: Result for 10 equally spaced phases $\varphi_2 = 2\pi n/(10 q)$
for $n = 0,1,\ldots, 9$ showing that for sufficiently large $T$, the
curve for $\varphi_2=0$ ($n=0$, blue) smoothly connects to the one for
$\varphi = 2\pi/q$.
Right column: The same but for $\varphi = 2\pi n/q$, $n=0, 1,\ldots,q-1$.
With increasing $T$, the curves start to coincide.
}
\label{fig:buildup}
\end{figure}

To underline this result, we also plot the curves within one $2\pi/q$
period and those for $\varphi_2$ equal to multiples of $2\pi/q$ separately,
but now for the $(q,p) = (7,5)$ resonance, see Figure~\ref{fig:buildup}.
The left column contains the results for values of $\varphi_2$ in the range
$[0,2\pi/q]$.  They show that only for a sufficiently large $T$, the
enveloping function is dominated by the $\ell=1$ component and resembles
the $\sinc$ conjectured in equation \eqref{asympbic}.  Moreover, the first
and the last curve smoothly connect to each other, which depicts how the
$2\pi/q$-periodicity emerges.  Accordingly, the curves for $\varphi_2$ at
multiples of $2\pi/q$ eventually coincide, as can be appreciated in the
right column of Fig.~\ref{fig:buildup}

As a remnant of finite propagation time $T$, we witness in all panels of
Figure~\ref{fig:buildup} an inclination of the resonance peak, which to a
smaller extent is noticeable also in Figure~\ref{fig:buildupall}.  It stems
from the global background of $\overline{\langle\sigma_z\rangle}_T$
visible in Figure~\ref{fig:overview}.  Owing to the scaling of the abscissa,
it diminishes with increasing $T$ and, in accordance with equation \eqref{asympbic},
it eventually vanishes.

\section{Conclusions}
\label{sec:conclusions}

We have studied the asymptotic limit of multichromatically driven,
dissipative dynamical systems.  It turned out that a strict distinction
between commensurable and incommensurable frequencies requires an infinite
propagation time.  Nevertheless, there exits a noticeable difference
between the two cases, namely that only for commensurable
frequencies the phases of the driving fields may matter. Moreover, the phase
dependence generally has a lower periodicity than the naively expected
$2\pi$.  While this implies non-generic features of the
resonances, the resulting peaks upon variation of the phase exhibit a
generic form given by sinc functions.

While the limiting behavior can be derived analytically, we have performed
numerical studies to see how the limits are approached.  For a
classical random walk on a lattice with bichromatically time-dependent
transition rates, the velocity has been obtained analytically up to the
numerical computation of a sum.  The response as a function of the
two driving frequencies shows how resonances emerge around rational values
of $\Omega_2/\Omega_1$.

The case of a dissipative two-level system has been treated fully
numerically. It revealed how with increasing propagation time, the generic
features of resonance peaks emerge, namely the sinc shape and the sub
$2\pi$ periodicity in the phase shift.  The width of the resonance peaks
at rational frequency quotients shrinks with increasing averaging time, such that
the background eventually appears flat and the peaks become pronounced.
The value of the response depends on the relative phase of the two
drivings, while in its vicinity, the response becomes phase independent.

An important point in practical calculations is that commensurable
frequency ratios with rather large numerator or denominator imply large
periods.  Then owing to the necessarily finite propagation time, this
periodicity may still not be manifest in the result.  In other words, up to
such finite time, the system behaves as if it were quasi-periodic, i.e., as
if the frequencies were incommensurable.  However, in particular for the
random-walk model, it may take even considerably longer until the peaks
assume their generic shape.
Quantitative statements about this issue still represent a challenge for
future investigations.

\begin{acknowledgement}
This work was supported by the Spanish Ministry of Science, Innovation, and
Universities through the CSIC Research Platform on Quantum Technologies
PTI-001 and via grants No.\ MAT2017-86717-P and FIS2017-86478-P.
\end{acknowledgement}

\bibliographystyle{epj}

\begin{thebibliography}{27}

\bibitem{Gammaitoni1998a}
L.~Gammaitoni, P.~H\"anggi, P.~Jung, F.~Marchesoni, Rev. Mod. Phys.
  \textbf{70}, 223 (1998)

\bibitem{CasadoPascual2003a}
J.~Casado-Pascual, J.~G\'omez-Ord\'o{\~n}ez, M.~Morillo, P.~H\"anggi, Phys.
  Rev. Lett. \textbf{91}, 210601 (2003)

\bibitem{anishchenko:2002}
V.~Anishchenko, A.~Neiman, A.~Astakhov, T.~Vadiavasova, L.~Schimansky-Geier,
  \emph{{Chaotic and Stochastic Processes in Dynamic Systems}} (Springer,
  Berlin, 2002)

\bibitem{freund:2003}
J.A. Freund, L.~Schimansky-Geier, P.~H{\"a}nggi, Chaos \textbf{13}, 225 (2003)

\bibitem{lindner:2004}
B.~Lindner, J.~Garcia-Ojalvo, A.~Neiman, L.~Schimansky-Geier, Phys. Rep.
  \textbf{392}, 321 (2004)

\bibitem{casado1:2005}
J.~Casado-Pascual, J.~G{\'o}mez-Ord{\'o}{\~n}ez, M.~Morillo, J.~Lehmann,
  I.~Goychuk, P.~H{\"a}nggi, Phys. Rev. E \textbf{71}, 011101 (2005)

\bibitem{Goychuk2006a}
I.~Goychuk, J.~Casado-Pascual, M.~Morillo, J.~Lehmann, P.~H\"anggi, Phys. Rev.
  Lett. \textbf{97}, 210601 (2006)

\bibitem{Reimann2002a}
P.~Reimann, Phys. Rep. \textbf{361}, 57 (2002)

\bibitem{Hanggi2009a}
P.~H\"anggi, F.~Marchesoni, Rev. Mod. Phys. \textbf{81}, 387 (2009)

\bibitem{CuberoRenzoni}
D.~Cubero, F.~Renzoni, \emph{Brownian Ratchets: From Statistical Physics to Bio
  and Nano-motors} (Cambridge University Press, Cambridge, 2016)

\bibitem{Flach2000a}
S.~Flach, O.~Yevtushenko, Y.~Zolotaryuk, Phys. Rev. Lett. \textbf{84}, 2358
  (2000)

\bibitem{Cubero2018a}
D.~Cubero, F.~Renzoni, Phys. Rev. E \textbf{97}, 062139 (2018)

\bibitem{Cubero2018b}
D.~Cubero, G.R. Robb, F.~Renzoni, Phys. Rev. Lett. \textbf{121}, 213904 (2018)

\bibitem{CasadoPascual2013a}
J.~Casado-Pascual, D.~Cubero, F.~Renzoni, Phys. Rev. E \textbf{88}, 062919
  (2013)

\bibitem{CasadoPascual2015a}
J.~Casado-Pascual, J.A. Cuesta, N.R. Quintero, R.~Alvarez-Nodarse, Phys. Rev. E
  \textbf{91}, 022905 (2015)

\bibitem{Reimann1997a}
P.~Reimann, M.~Grifoni, P.~H\"anggi, Phys. Rev. Lett. \textbf{79}, 10 (1997)

\bibitem{Lehmann2003b}
J.~Lehmann, S.~Kohler, P.~H\"anggi, A.~Nitzan, J. Chem. Phys. \textbf{118},
  3283 (2003)

\bibitem{Kohler2005a}
S.~Kohler, J.~Lehmann, P.~H\"anggi, Phys. Rep. \textbf{406}, 379 (2005)

\bibitem{Forster2015b}
F.~Forster, M.~M\"uhlbacher, R.~Blattmann, D.~Schuh, W.~Wegscheider, S.~Ludwig,
  S.~Kohler, Phys. Rev. B \textbf{92}, 245422 (2015)

\bibitem{Platonov2015a}
S.~Platonov, B.~K\"astner, H.W. Schumacher, S.~Kohler, S.~Ludwig, Phys. Rev.
  Lett. \textbf{115}, 106801 (2015)

\bibitem{Morito1979a}
S.~Morito, H.M. Salkin, Fibonacci Quart. \textbf{17}, 361 (1979)

\bibitem{Morito1980a}
S.~Morito, H.M. Salkin, Acta Inform. \textbf{13}, 379 (1980)

\bibitem{vanKampen1992a}
N.G. van Kampen, \emph{Stochastic processes in physics and chemistry}
  (North-Holland, Amsterdam, 1992)

\bibitem{Hanggi1990a}
P.~H\"anggi, P.~Talkner, M.~Borkovec, Rev. Mod. Phys. \textbf{62}, 251 (1990)

\bibitem{Kohler1997a}
S.~Kohler, T.~Dittrich, P.~H\"anggi, Phys. Rev. E \textbf{55}, 300 (1997)

\bibitem{Grifoni1998a}
M.~Grifoni, P.~H\"anggi, Phys. Rep. \textbf{304}, 229 (1998)

\bibitem{Breuer2003a}
H.P. Breuer, F.~Petruccione, \emph{Theory of Open Quantum Systems} (Oxford
  University Press, Oxford, 2003)

\end{thebibliography}

\end{document}